\documentclass[conference]{IEEEtran}
\IEEEoverridecommandlockouts
\usepackage{cite}
\usepackage{amsmath,amssymb,amsfonts}
\usepackage{algorithmic}
\usepackage{graphicx}
\usepackage{textcomp}
\usepackage{xcolor}
\def\BibTeX{{\rm B\kern-.05em{\sc i\kern-.025em b}\kern-.08em
    T\kern-.1667em\lower.7ex\hbox{E}\kern-.125emX}}
\usepackage{subcaption}
\usepackage{makecell}
\usepackage{multirow}
\usepackage[linesnumbered,ruled,vlined]{algorithm2e}
\usepackage{amsmath}
\usepackage{pifont}

\usepackage{url}
\newcommand{\cmark}{{\ding{51}}}%
\begin{document}

\title{PRISM: Privacy Preserving Healthcare Internet of Things Security Management}

    \author{\IEEEauthorblockN{Savvas Hadjixenophontos}
    \IEEEauthorblockA{
    \textit{Imperial College London}\\
    }
    \and
    \IEEEauthorblockN{Anna Maria Mandalari}
    \IEEEauthorblockA{\textit{University College London} \\
    }
    \and
    \IEEEauthorblockN{Yuchen Zhao}
    \IEEEauthorblockA{\textit{University of York} \\
    }
    \and
    \IEEEauthorblockN{Hamed Haddadi}
    \IEEEauthorblockA{\textit{Imperial College London} \\
    }
   
    }

\maketitle

\begin{abstract}
Consumer healthcare Internet of Things (IoT) devices are gaining popularity in our homes and hospitals. These devices provide continuous monitoring at a low cost and can be used to augment high-precision medical equipment. However, major challenges remain in applying pre-trained global models for anomaly detection on smart health monitoring, for a diverse set of individuals that they provide care for.
In this paper, we propose PRISM, an edge-based system for experimenting with in-home smart healthcare devices.  
We develop a rigorous methodology that relies on automated IoT experimentation. We use a rich real-world dataset from in-home patient monitoring from 44 households of People Living With Dementia (PLWD) over two years. Our results indicate that anomalies can be identified with accuracy up to 99\% and mean training times as low as 0.88 seconds. While all models achieve high accuracy when trained on the same patient, their accuracy degrades  when evaluated on different patients.
\end{abstract}
\section{Introduction}

Internet of Things (IoT) devices are increasingly being used in the healthcare industry~\cite{Habibzadeh2020, kashani2021systematic}, providing patient monitoring and in-home healthcare solutions for different groups such as the elderly and People Living With Dementia (PLWD)~\cite{ ray2019systematic}. These devices often come with a range of sensors and require access to a number of sources of personal data and continuous internet connectivity. Traditionally, the personal data which is highly private and sensitive is sent to the cloud for applications such as training machine learning (ML) models and using the models to generate inferences~\cite{Jaiswal2017}. Such a centralized system structure poses significant privacy and security issues to data subjects, especially for in-home healthcare where data is continuous and personal~\cite{Meingast}.

Edge-based systems, which provide a decentralized paradigm, have recently become a preferable choice for designing and implementing IoT applications, especially in scenarios in which privacy is critical~\cite{mandalari2021blocking}. Instead of sending data to the cloud, an IoT application at the edge keeps personal data locally on edge devices. It can use the data to train ML models such as Deep Neural Networks (DNNs)~\cite{velasco2018optimum} and generate inferences from the models locally, thereby protecting the privacy of data subjects~\cite{zhao2018privacy,osia2018private}. 

Although existing research has shown that edge-based IoT applications have acceptable overall performance~\cite{pace2018edge}, the accuracy of the applications at the individual level may vary from one patient to another due to the difference in their data~\cite{wu2021edge}.
Models developed for lab-based scenarios do not always translate to complex scenarios with patient monitoring (due to the presence of carers, family members, or unusual daily activity patterns). Moreover, experiments indicate that while models can accurately identify devices and anomalies, they also rapidly decay over time, indicating the need for continuous, local, and personalized retraining~\cite{kolcunrevisiting}. 

Previous work has resorted to machine learning to analyze anomalies in PLWD~\cite{palermo2021designing}.  The usual approach entails training machine learning models offline or in a cloud environment. 
However, the training and validation of these models is
done on a particular patient, and for a limited time period.
Different patients have different usage patterns, and their behavior might evolve over time. For example, a patient might
have a certain sleeping routine and consequently different interaction
with their IoT devices, but interact differently with the devices while
he is on medications. 
It is therefore important to understand whether adopting ML in edge-based systems benefits all individual households. We examine this assumption at the individual level in real healthcare settings and want to understand how different aspects of data affect accuracy at the edge for anomaly detection.

In this paper, we propose PRISM, an edge-based system for in-home smart healthcare anomaly detection. We characterize the individual-level performance of edge activity inference model in terms of the trade-off between inference time and accuracy, and we discuss applications and future directions. We achieve this by mixing real data collected in an extensive clinical study with anomaly injection at the newtwork gateway. We then focus on training at the edge, where compute capability is limited. 
Based on our comparison of different models, we choose neural network-based models for edge deployment.
PRISM has the ability to deploy a number of privacy-preserving, dynamically configurable DNN models for local training, and inference at the edge. 
We perform a total of 190,980 rigorous automated controlled experiments, and we leverage a dataset composed of 44 households including 22 healthcare IoT devices.

The main contributions of the paper are as follows:
\begin{itemize}
\item We develop a system for injecting realistic anomalies for healthcare IoT devices; 
\item We compare three different types of anomalies and show that in all cases the accuracy changes with the train window size, with accuracy up to 99\%;
\item We show that models need to be updated using data specific to the patient. A model updated using data from one
patient does not perform well on another patient and vice versa;
\item We demonstrate that training the model at the edge of the network on a representative edge device (Raspberry Pi) is feasible, with  training times as low as 0.88 seconds.
\end{itemize}

The remainder of the paper is organized as follows.  We provide an overview of our system (Section~\ref{system}). We describe our data set, testbed, and the experiments conducted (Section~\ref{experiment}). In Section~\ref{results} we show the anomaly detection accuracy and we demonstrate how models can be trained at the edge. We then discuss limitations and future challenges (Section~\ref{discussion}). Finally, We provide an overview of related work (Section~\ref{related}) and conclude our work (Section~\ref{conclusion}).

We make our code and anomaly dataset available. \footnote{\url{https://github.com/IOTPRISM/prism}}
\section{System Design}
\label{system}

\begin{figure}[t]
    \centering
    \includegraphics[scale = 0.38]{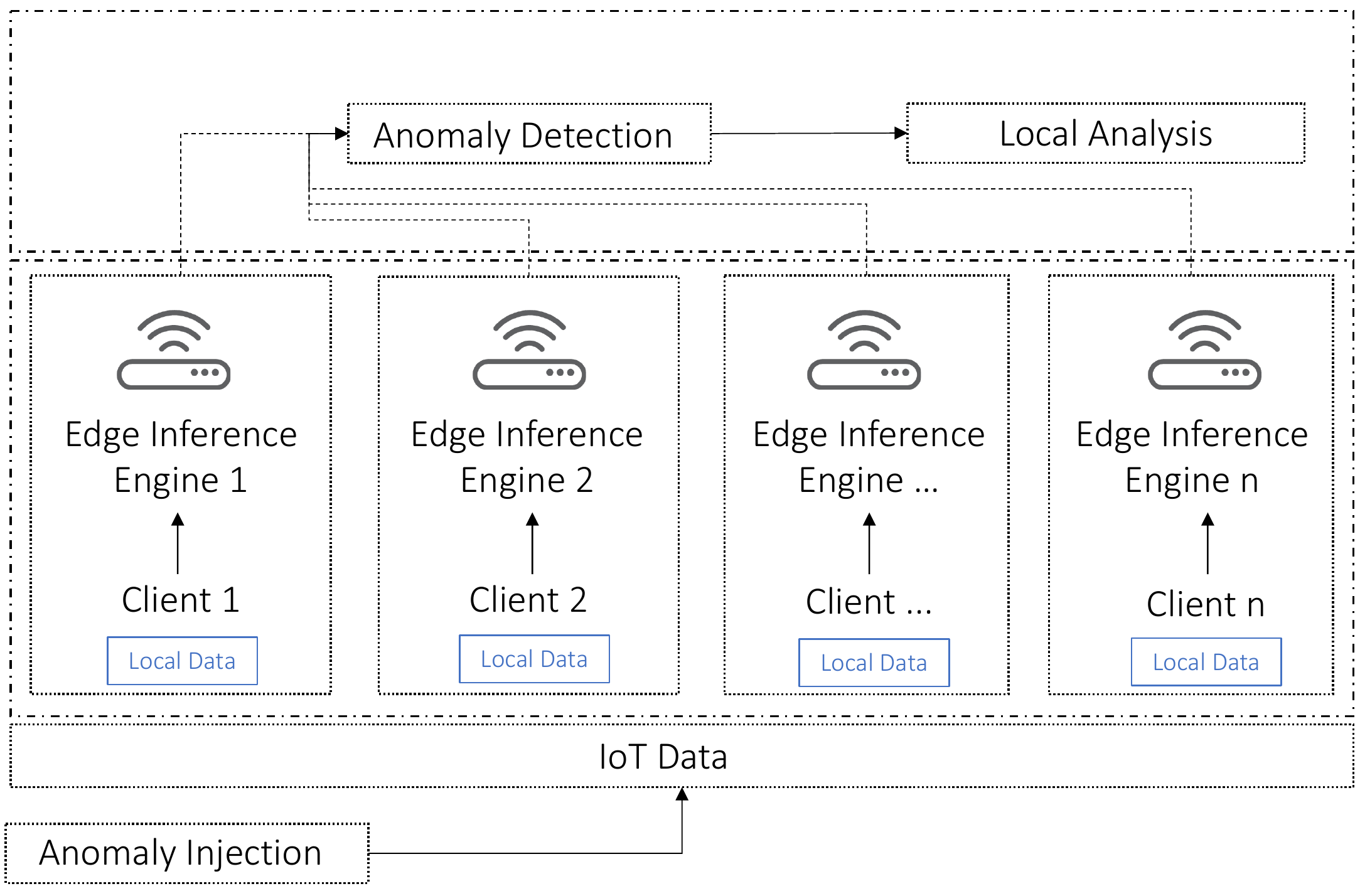}
    \caption{System overview of PRISM. Each edge device (i.e., home gateway) keeps its IoT data locally for ML inference and training. Anomaly data is injected for each IoT device and the anomaly detection mechanism is applied and evaluated individually.}
    \label{fig:system_overview}
\end{figure}

In this section, we report the system design for PRISM.
We consider anomaly detection of IoT device data as the inference task and examine the performance of PRISM. We simulate the anomaly injection to real patients' data and evaluate the trade-off between anomaly detection time and accuracy among them. Figure~\ref{fig:system_overview} shows an overview of our system. PRISM consists of two components: (i) \emph{anomaly injection module} and (ii) \emph{anomaly detection module}. The description and some examples of anomalies are reported in the next Section.


\subsection{Anomaly Injection Module}
The anomaly injection module allows to select the type of anomaly to be injected for experimenting with the IoT data. 
The module allows selection of:
(i) the number of anomalies to be injected in the data frame;
(ii) the type of anomaly;
(iii) the data frame in which the anomalies will be injected.

The module reads the earliest and latest timestamp of the IoT readings and then creates a random timestamp within that range in which the anomaly will be inserted at. 
At the end of the process, the original data frame is concatenated with the anomaly data frame. The concatenated result is then reordered based on the timestamp of the combined readings.

\subsection{Anomaly Detection Module}
\label{alpha}
The anomaly detection module is based on Neural Network (NN). 
We choose neural networks for the predictive model as they can learn complex behavior. It is also possible to update a neural network's weights when more data becomes available~\cite{chen1999rapid} and different configurations of the network can be readily compared~\cite{hunter2012selection}. 
We implement an unsupervised learning system, hence we do not label the anomalous differently from the non-anomalous data. 

The anomalies are detected by comparing the loss of the current validation batch with a threshold based on the loss generated during training. The threshold is calculated based on the average training loss$^*$ of all batches during the last epoch, multiplied with a coefficient $\alpha$ which is the tuning parameter of the threshold. The linear equation corresponding to the threshold computation is:
$$Threshold = \alpha \times mean(Training\;Loss^*)$$
If the loss calculated exceeds that threshold, the batch is classified as anomalous.

The module parameters are initialized during training. We use the Stochastic Gradient Descent (SGD)~\cite{moulines2011non} as our optimization function whose hyperparameters are set as $3e^-5$ (learning rate), and the batch size and number of epochs are initialized with the values of 1 and 100 respectively. The permutations which are initialized are used for the batch size implementation during training.

The output anomalies are then tested to see which ones are correctly identified, and which ones are false positives.
We calculate the accuracy using the following equation\small : \\ $$ Accuracy = \frac{True\; pos \;+ \;True\;neg}{True\; pos +True\;neg+False\;pos+False\;neg}$$\normalsize
Each training and validation time window combination is repeated 30 times for each IoT device. Note that during each repetition the point in time from which the time windows are extracted changes. 

\section{Experimental Setting}
\label{experiment}

In this section we describe our experimental setting, the dataset we use and how we define and inject the anomalies.

As a representative edge device, we use a Raspberry Pi model 4
with 4 GB of RAM. The Raspberry Pi is running the Ubuntu 22.04
operating system with Python 3.9.7, on which a traditional ML stack
is installed (numpy, scipy, pandas, scikit-learn, and PyTorch).

\subsection{Dataset}

We leverage the data collected by the UK Dementia Research Institute Care Research and Technology Centre (UK DRI CR\&T)~\cite{palermo2021designing}, which contains the in-home monitoring IoT data that includes motion sensors, physiological measurements, and the use of kitchen appliances from 44 homes of  PLWD, between April 2019 and June 2021. There are 22 different IoT devices installed in the household of each patient.
Table~\ref{tab:devices} shows the IoT devices installed in each home, the function they provide, the data format, and whether the data collection is continuous. 

\begin{table}[h!]\scriptsize
    \centering
    \begin{tabular}{|c|c|c|c|}
    \hline
    \textbf{Function}&\textbf{Format}&\textbf{IoT Device}&\textbf{Continuous}\\\hline
   \textbf{Location}  & Binary & WC, bathroom, bedroom, corridor& \\&&dining room, hallway&-\\
   &&kitchen, living room, lounge&\\&& office, study& \\
\hline
    \textbf{Door}  & Binary& back door, conservatory&\\ &&fridge door, front door&\cmark \\
   &&garage, main door&\\ && secondary, utility&\\
   \hline
    \textbf{Appliances}  & Binary & iron use, kettle use, microwave use&-  \\
   &&socket use, toaster use&\\
    \hline
    \textbf{Temperature}  & Float & temperature, body temperature&\\&& skin temperature&\cmark  \\
      \hline
    \textbf{Health Related}  & Float & blood pressure, body mass index&\\&& body muscle mass, body weight  &-  \\
   &&heart rate, body fat&\\ &&body water, bone mass&\\
     \hline
   \textbf{Light}  & Integer & light level &\cmark  \\
      \hline
    \textbf{Sleep Event}  & Binary & sleep event, sleep mat snoring &  \\
   &Float & sleep mat heart rate&\cmark\\&& sleep mat respiratory rate &  \\
      &Integer & sleep mat state, agitation &-  \\
     \hline
    \end{tabular}
        \vspace{0.5cm}
    \caption{The corresponding functions of the IoT devices, the data format and whether they are continuously reported.}
    \label{tab:devices}
\end{table}

\subsection{Data Pre-processing}
The data pre-processing part is done after the anomaly injection module. Since several patients are used for training and validation, the resulting data frames of each patient are concatenated together to create the training data and the validation data which are then individually broken down to inputs and labels being X\_trainD / X\_valD and Y\_trainD / Y\_valD respectively. Finally, the data frames are converted into PyTorch Tensors, and then into floats, resulting in the final training and validation data (which are used by the machine learning model). 

Each dataset is associated with a time delta value t between
0.25 and 24 hours and is filtered to contain only the values whose
timestamps fall between the time delta.

\subsubsection{Patient Activity Pattern}  
By studying the data activity of the patients, we try to detect whether the patients have behavioral patterns and therefore understand whether the in-home monitoring data can be used to build a model. 
By analyzing the activity of each patient, we note that each patient has a behavioral pattern, which can be learned by a DNN module. 

\subsubsection{IoT Device Selection}
The IoT devices we adopt for anomaly detection are reported in Table~\ref{tab:devices}.
We remove the IoT devices that have a rare reading (i.e., Body Mass Index, Body Muscle Mass) or the devices only applied to a minority group of patients (Blood Pressure, body temperature, body weight, body mass index, body muscle mass, total Body Fat, Total Body Water, and Total Bone Mass).

\subsubsection{Data Pre-processing} Initially, the data from an individual patient is loaded to our model. During this stage, an additional feature is added to the data frame, which captures the difference in time between consecutive readings. This feature is added to provide the frequency of readings. Hence, the input features are the readings, and the difference in time. 
We split the data into training and testing data.
The size of the training and testing data depends on the size of the time window used for training and validation. We use six different time windows: 24, 3 hours, and 15 minutes. 

\subsection{Anomaly Definition}
In this section, we describe the anomalies injected. We consider three types of anomalies, each adapted to the kind of IoT devices we have in the data set. 

\noindent \textbf{On-Off.}
The On-Off anomaly refers to injecting abnormal frequency of on/off status to the binary value-based IoT devices. For example, the anomaly corresponds to the door sensor continuously closing and opening at an abnormal amount and rate. 
We manually create on/off events and inject them into the raw data.

\noindent \textbf{Variance.} 
Time series variance anomalies refer to the injection of values resembling abnormal variations centered around the standard readings. The variance anomalies follow the Gaussian distribution.  

\noindent \textbf{Spike.} 
Time series spikes refer to unexpectedly high values, it is well characterized to investigate the effect of an exposure spike on an outcome variable. We consider a spike to be an increase in the data series followed by an immediate return to the underlying level of the data series.
We inject a fixed number of spikes in random periods.

\begin{figure*}[t!]
     \centering
     \begin{subfigure}[t]{0.4\textwidth}
         \centering
         \includegraphics[width=\textwidth]{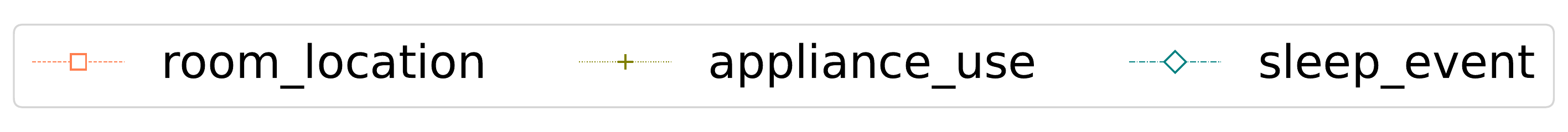}
     \end{subfigure}
     
     \begin{subfigure}[t]{0.275\textwidth}
         \centering
         \includegraphics[width=\textwidth]{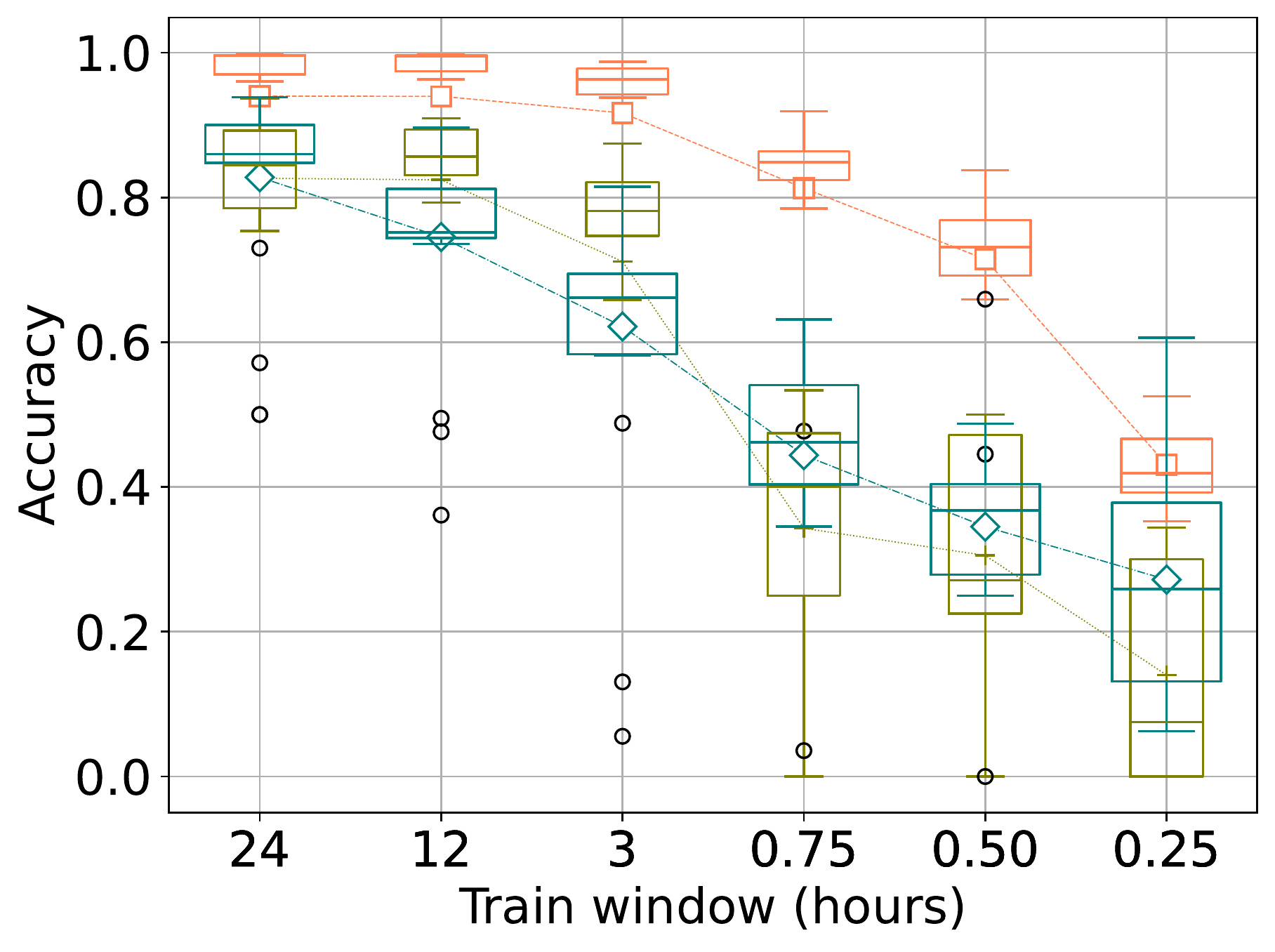}
         \caption{24h}
         \label{fig:Validation Window 24h}
     \end{subfigure}
     \begin{subfigure}[t]{0.275\textwidth}
         \centering
         \includegraphics[width=\textwidth]{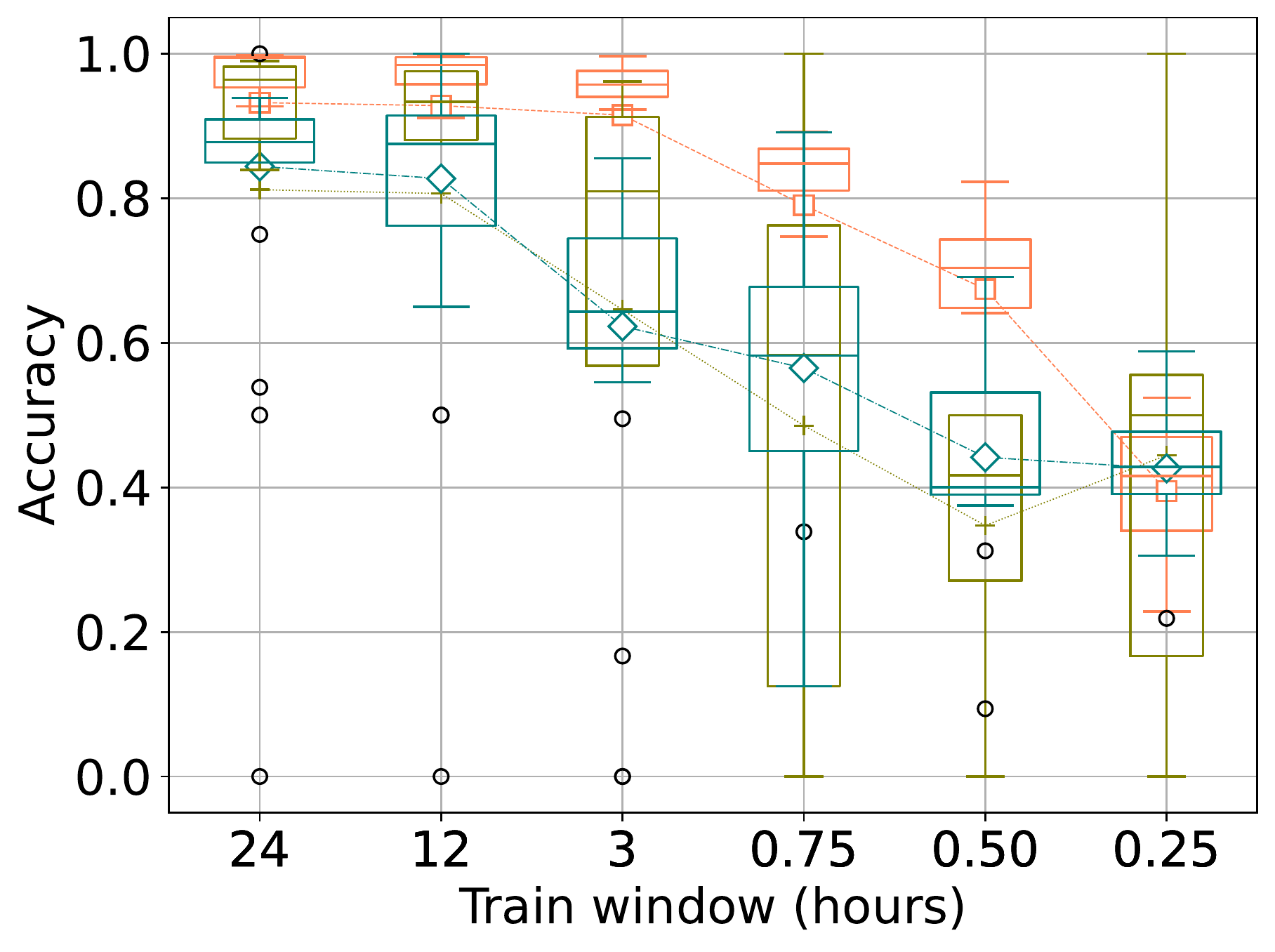}
         \caption{3h}
         \label{fig:Validation Window 3h}
    \end{subfigure}
    \begin{subfigure}[t]{0.275\textwidth}
         \centering
         \includegraphics[width=\textwidth]{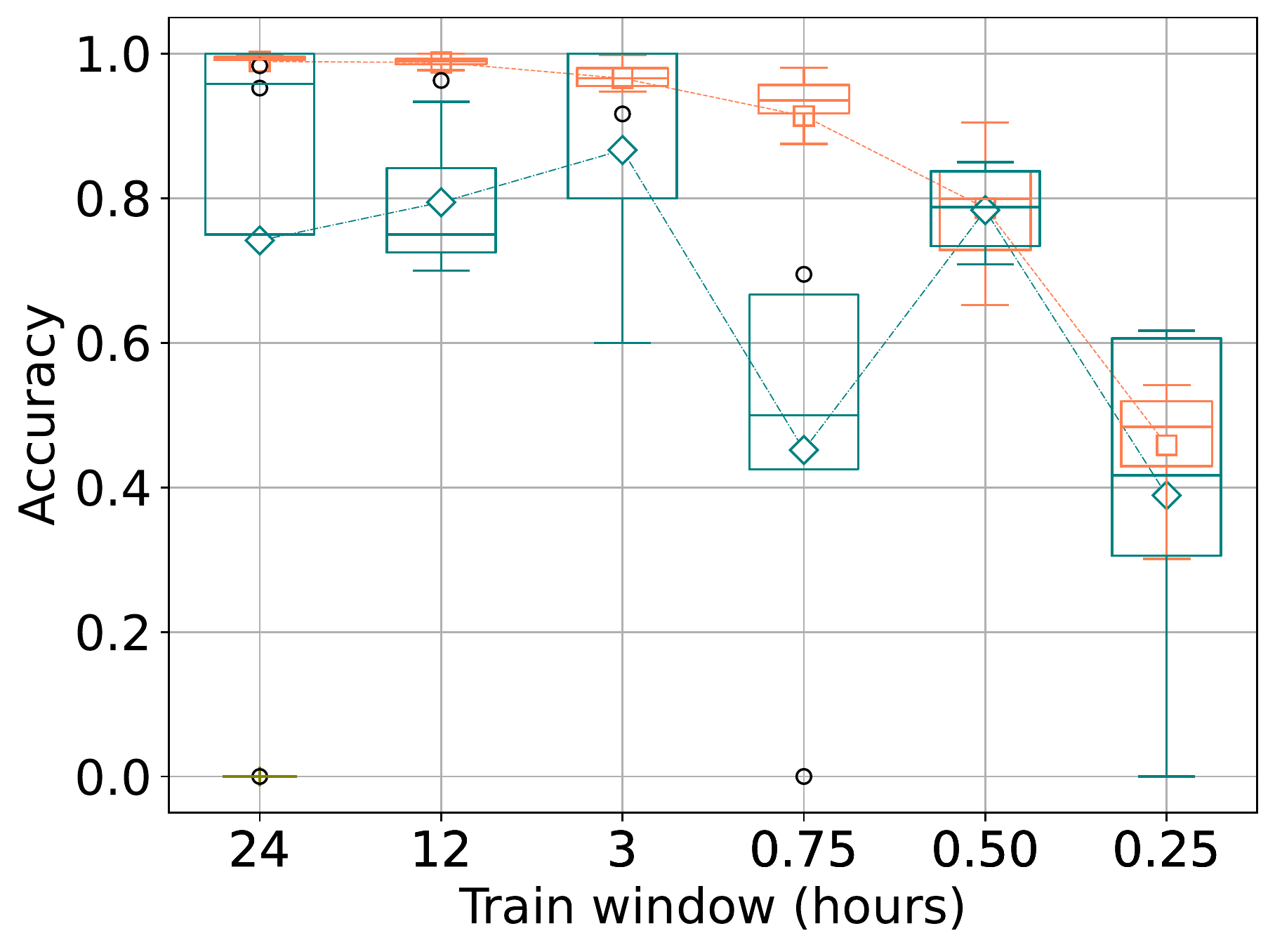}
         \caption{15min}
         \label{fig:Validation Window 15min}
    \end{subfigure}
    \caption{On-Off Anomaly. The anomaly detection accuracy changes with training window size and different validation window sizes. Each plot shows a validation window size (24h, 3h, 15min). The box plot shows the detection performance spread based on patients' data.}
    \label{fig:creat_anomaly}
\end{figure*}

\begin{figure*}
     \centering
     
     \begin{subfigure}[t]{0.45\textwidth}
         \centering
         \includegraphics[width=\textwidth]{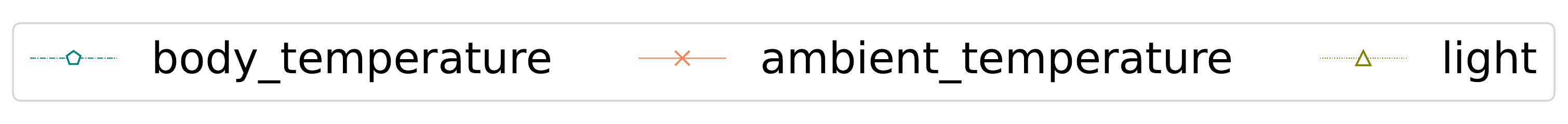}
     \end{subfigure}
     
     \begin{subfigure}[t]{0.275\textwidth}
         \centering
         \includegraphics[width=\textwidth]{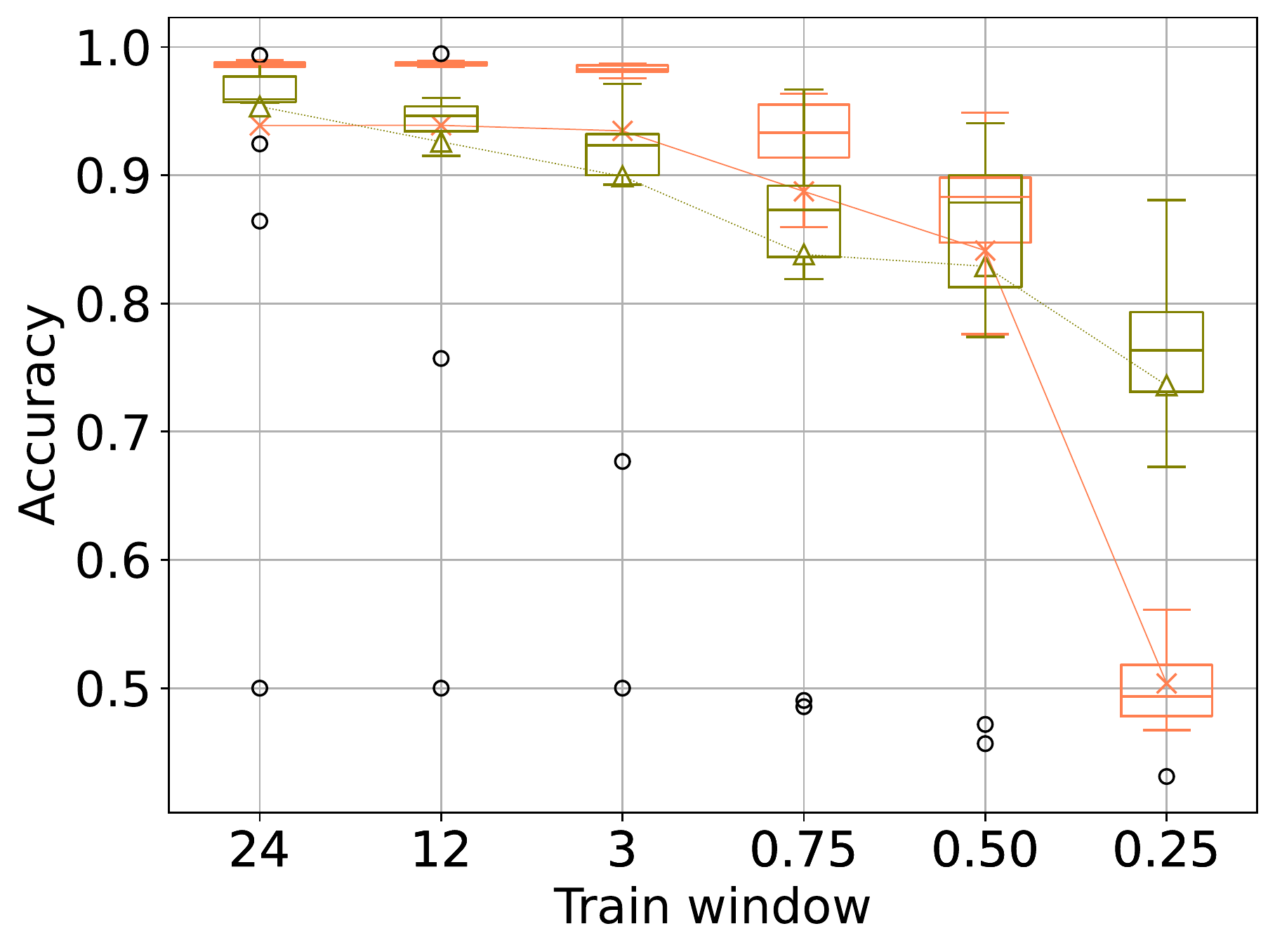}
         \caption{24h}
         \label{fig:Validation Window 24h}
     \end{subfigure}
     \begin{subfigure}[t]{0.275\textwidth}
         \centering
         \includegraphics[width=\textwidth]{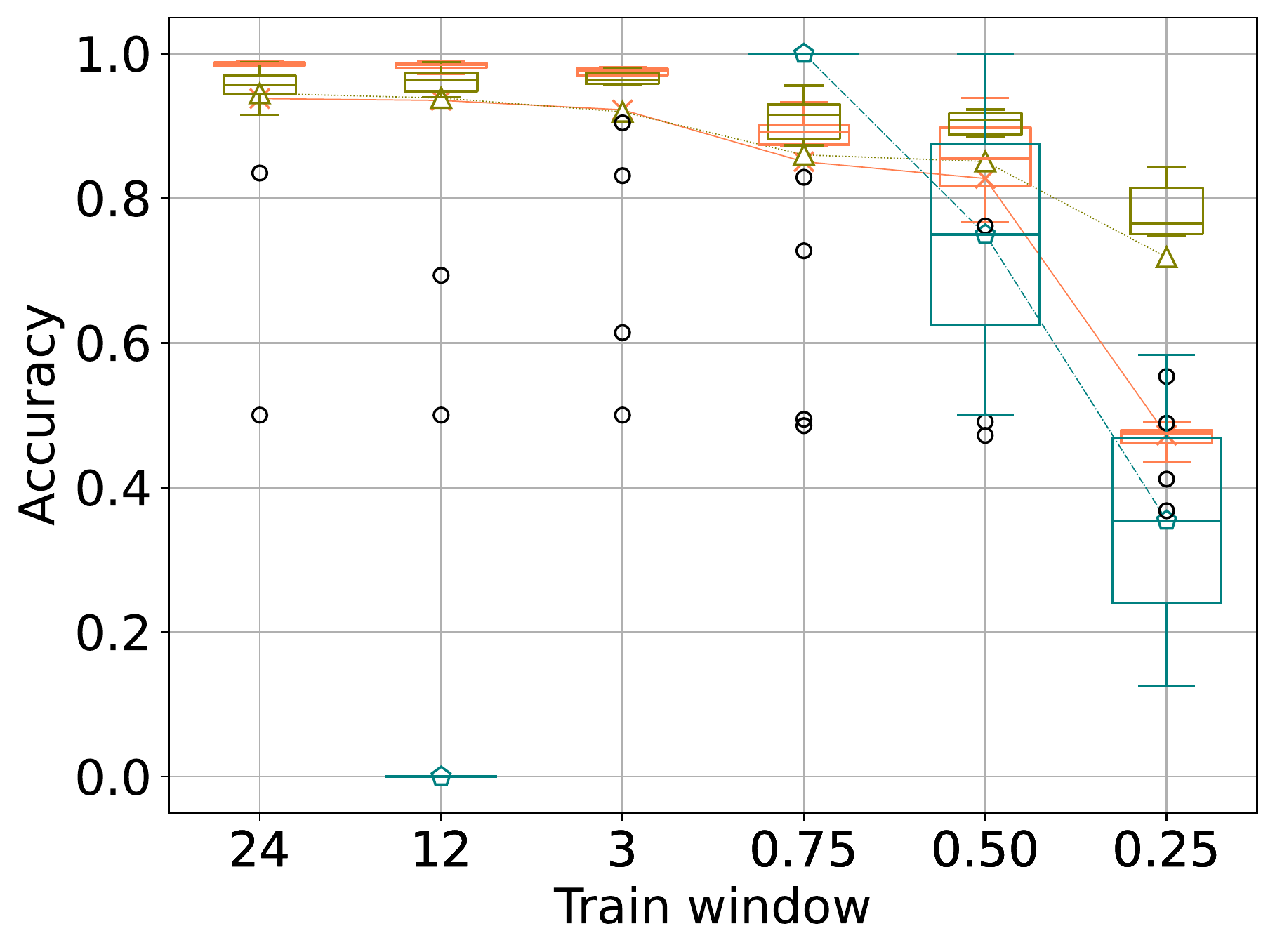}
         \caption{3h}
         \label{fig:Validation Window 3h}
    \end{subfigure}
    \begin{subfigure}[t]{0.275\textwidth}
         \centering
         \includegraphics[width=\textwidth]{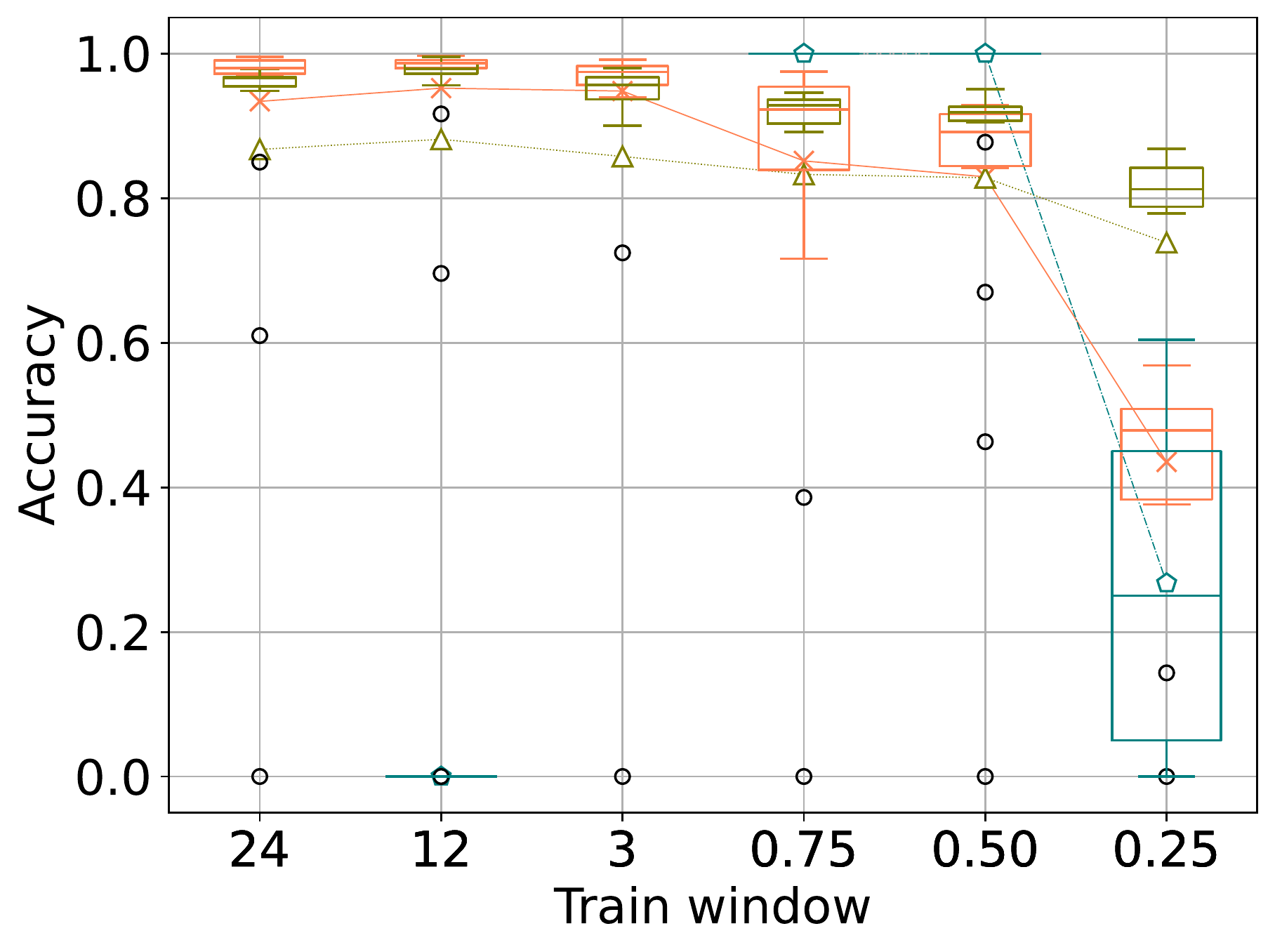}
         \caption{15min}
         \label{fig:Validation Window 15min}
    \end{subfigure}
    \caption{Variance Anomaly. The anomaly detection accuracy changes with training window size and different validation window sizes. Each plot shows a validation window size (24h, 3h, 15min). The box plot shows the detection performance spread based on patients' data.}
    \label{fig:variance_anomaly}
\end{figure*}

\begin{figure*}
     \centering
     \begin{subfigure}[t]{0.45\textwidth}
         \centering
         \includegraphics[width=\textwidth]{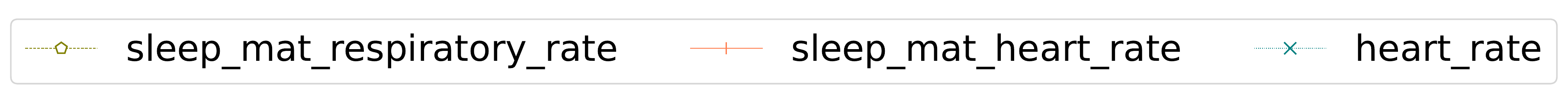}
     \end{subfigure}
     
     \begin{subfigure}[t]{0.275\textwidth}
         \centering
         \includegraphics[width=\textwidth]{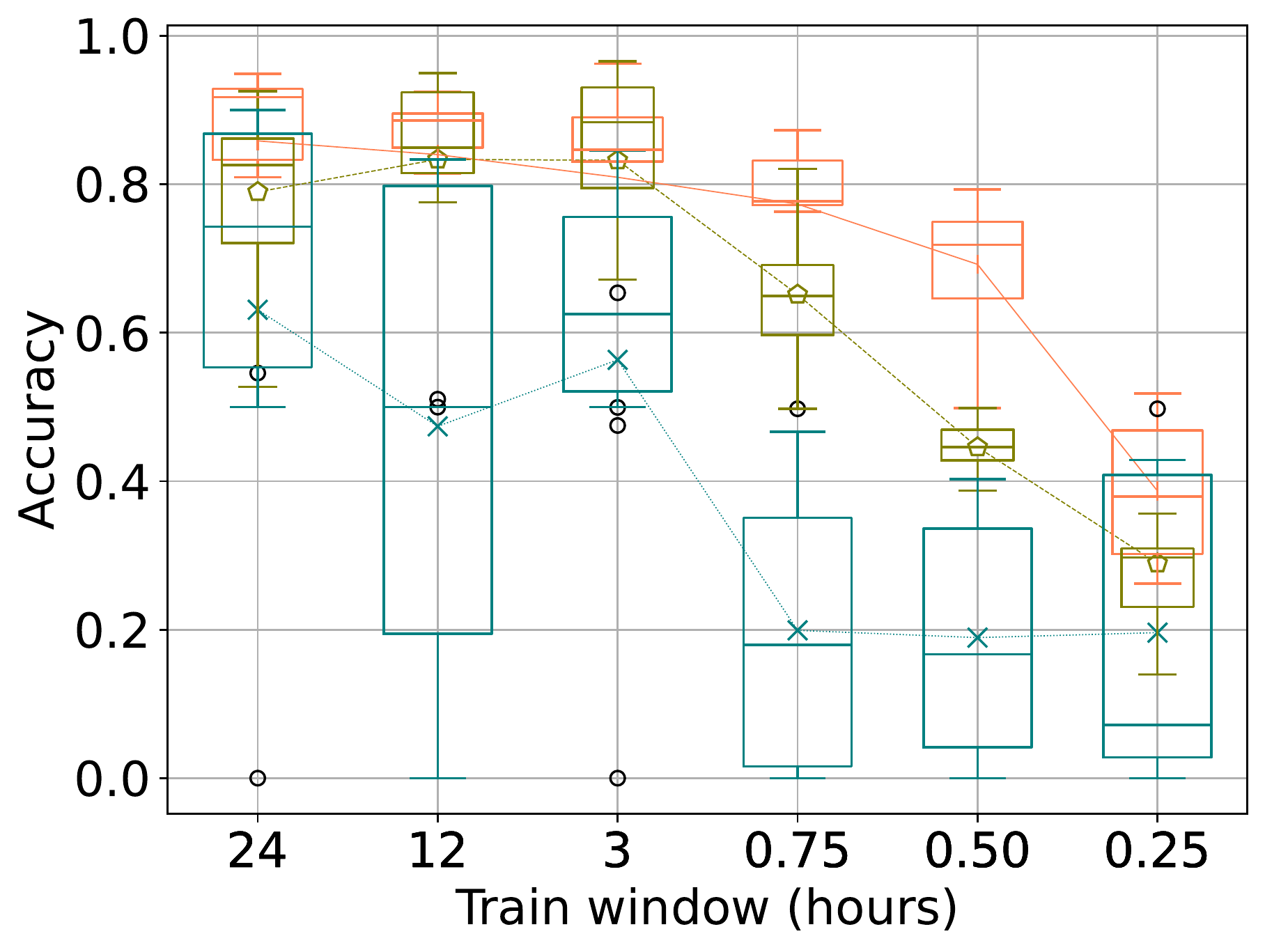}
         \caption{24h}
         \label{fig:Validation Window 24h}
     \end{subfigure}     
    \begin{subfigure}[t]{0.275\textwidth}
         \centering
         \includegraphics[width=\textwidth]{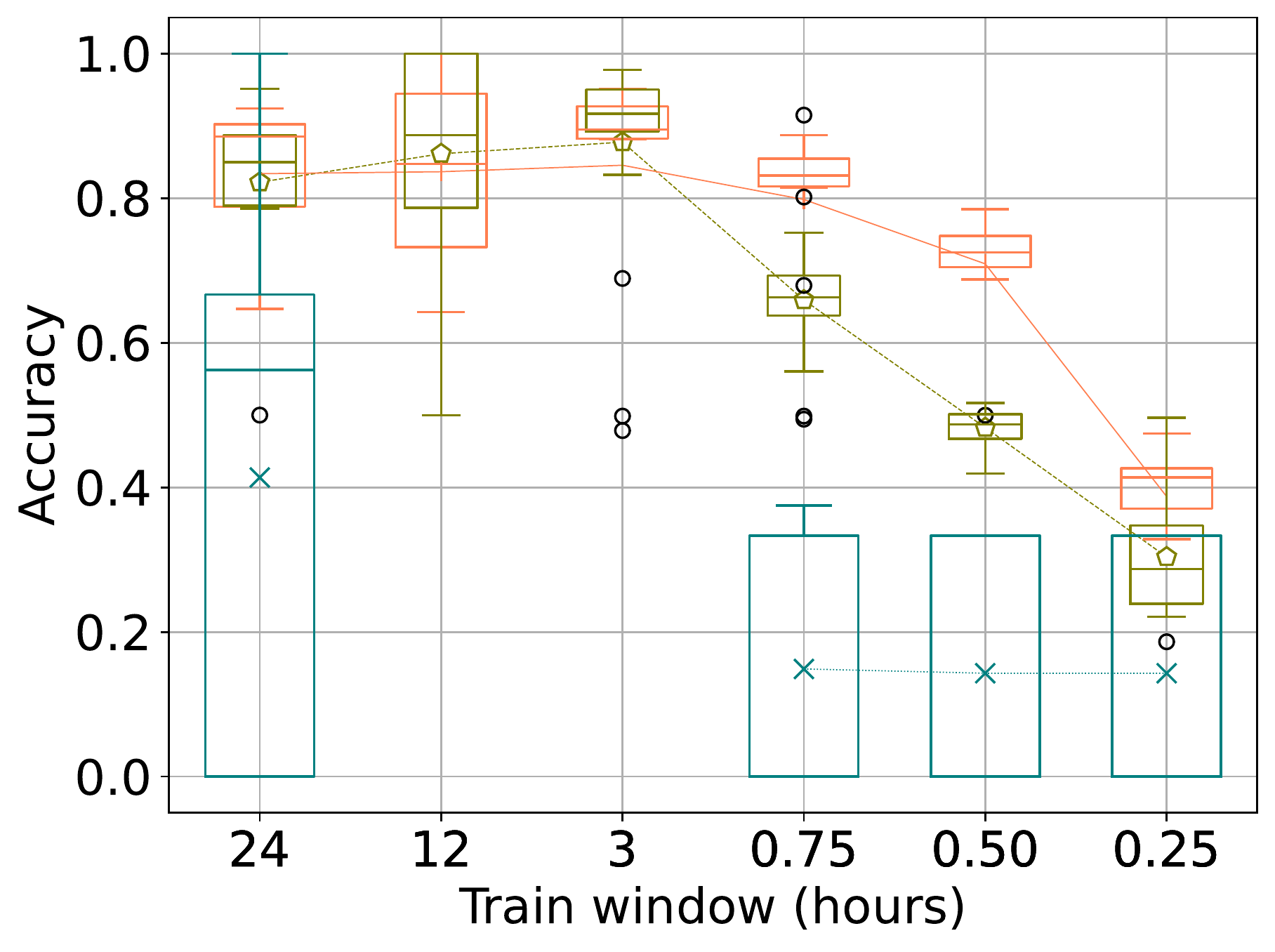}
         \caption{3h}
         \label{fig:Validation Window 3h}
    \end{subfigure}
    \begin{subfigure}[t]{0.275\textwidth}
         \centering
         \includegraphics[width=\textwidth]{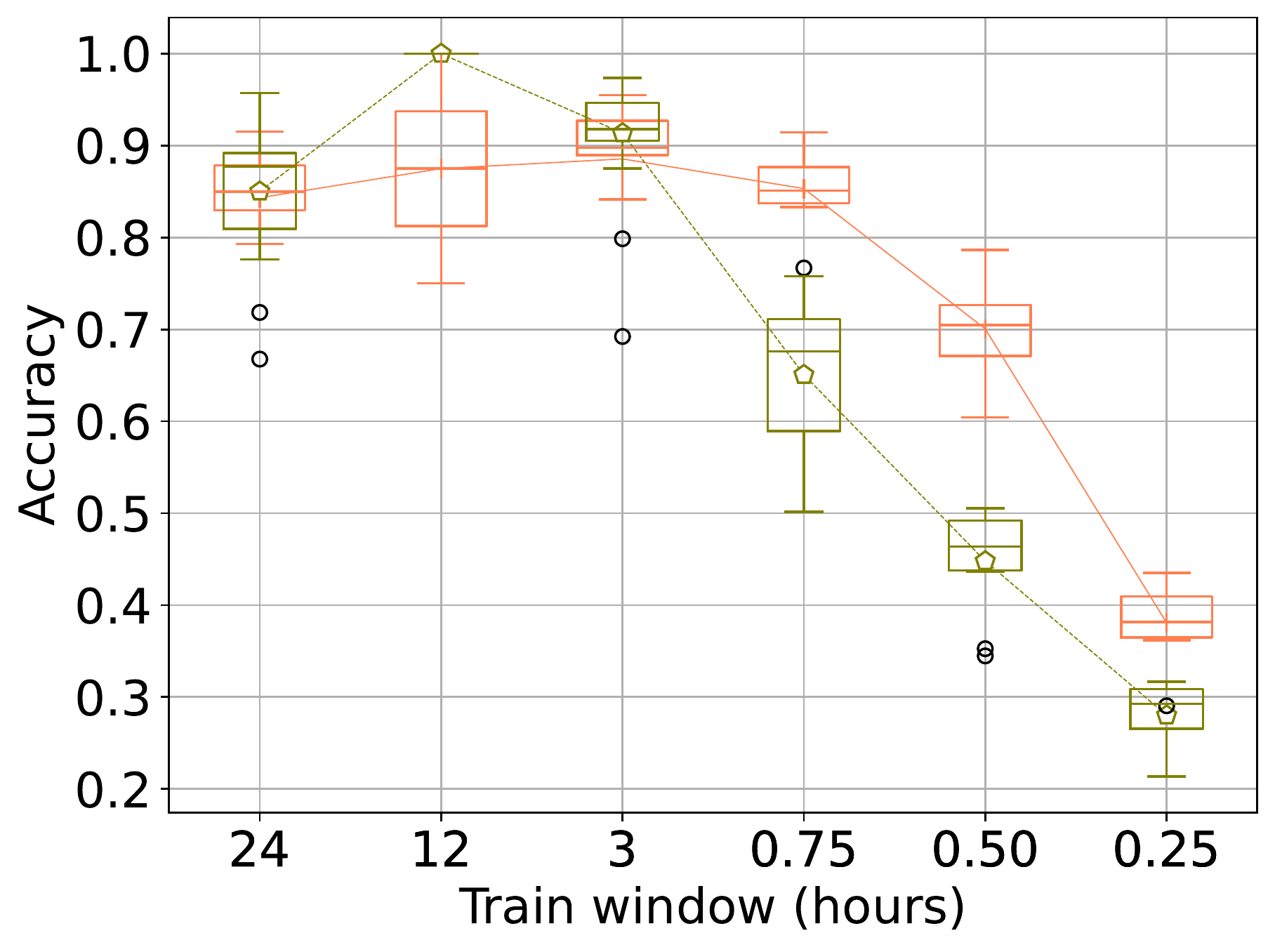}
         \caption{15min}
         \label{fig:Validation Window 15min}
    \end{subfigure}
    \caption{Spike Anomaly. The anomaly detection accuracy changes with training window size and different validation window sizes. Each plot shows a validation window size (24h, 3h, 15min). The box plot shows the detection performance spread based on patients' data.}
    \label{fig:spike_anomaly}
\end{figure*}

\section{Evaluation}
\label{results}

In this section, we discuss the performance of PRISM. We characterize the threshold for anomaly detection and the individual-level performance of edge activity inference model in terms of the trade-off between inference time and accuracy. We also determine the average training and inference running times at the edge.

\noindent \textbf{Anomaly Detection Accuracy.}
We first discuss the accuracy of the edge activity inference model when applied to different patients' data.

Figure~\ref{fig:creat_anomaly}, Figure~\ref{fig:variance_anomaly}, and Figure~\ref{fig:spike_anomaly}  show the average accuracy of different patients' data with three anomaly types and different training and validation windows. 
Unsurprisingly, a decrease in the training time window results in a decrease in accuracy. This is because a smaller time window results in fewer data used when training the model. 
IoT devices with larger numbers of data (i.e., Room Location) have smaller deviations and more stable performance. The validation time window does not affect the accuracy of the model. 

\begin{figure}[t]
         \centering
         \includegraphics[scale=0.45]{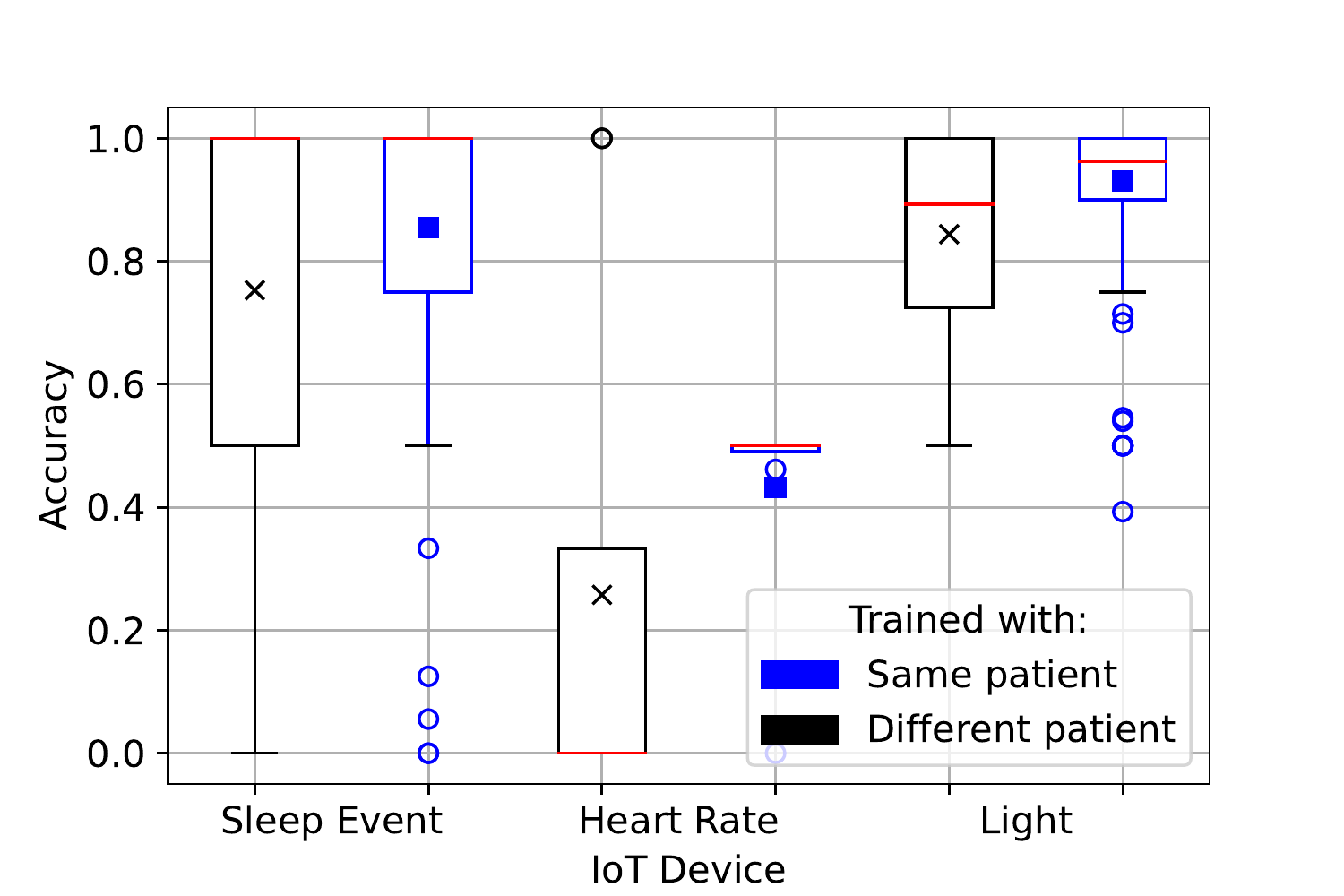}
         \caption{Average accuracy across all patients while training and validating with the same and different patients.}
         \label{fig:Trainingbox}
\end{figure}

\begin{figure}[t]
         \centering
         \includegraphics[scale=0.53]{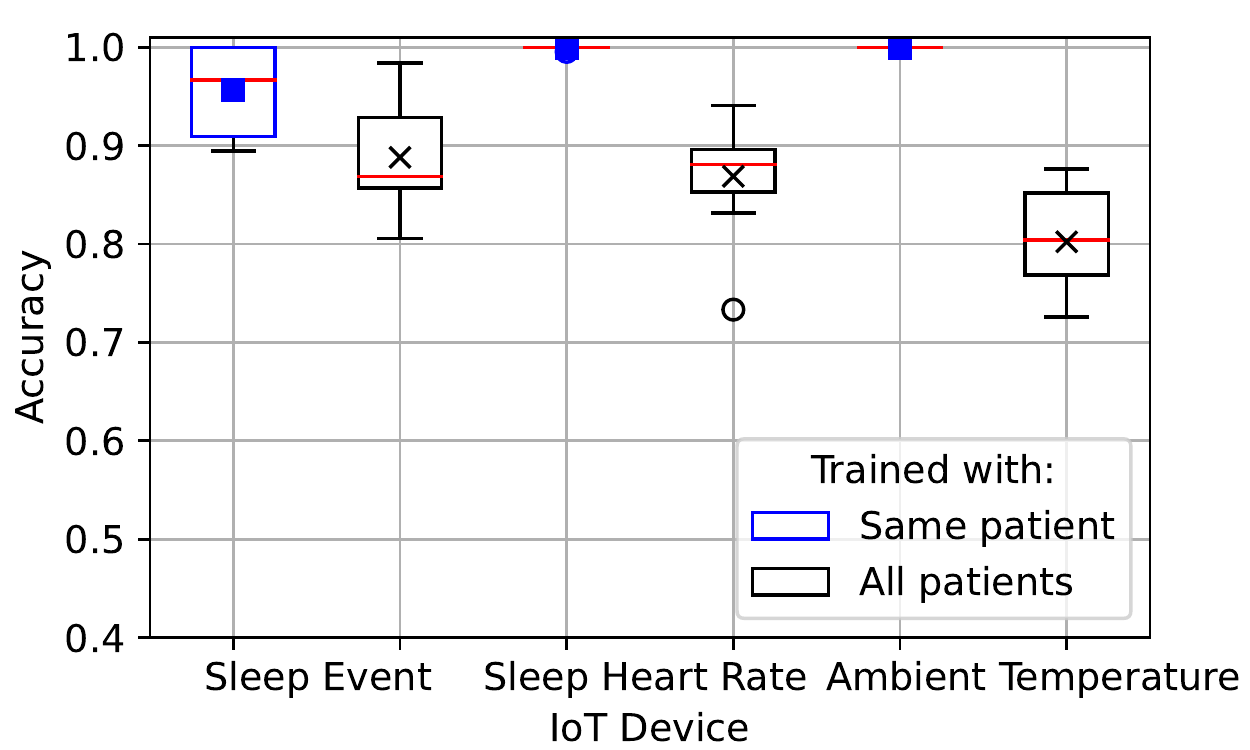}
         \caption{Average accuracy across all patients while training with all patients and validating with one patient, compared to training with all and validating with one patient.}
         \label{fig:all_train}
\end{figure}

\noindent \textbf{Personalized Models.}
Figure~\ref{fig:Trainingbox} shows the average accuracy  across all patients while training and validating with the same and different
patients. A model updated using data from one patient does not perform well on another patient and vice versa.
Figure~\ref{fig:all_train} shows how the model accuracy is affected when training the model with data from all patients and then validating using one patient, for three different IoT devices. The accuracy decreases when training the model with all patients. This shows that a model updated with data specific to each patient will achieve better performance, highlighting the need for solutions at the edge and using personalized models.

\begin{figure}
    \centering
        \includegraphics[scale=0.45]{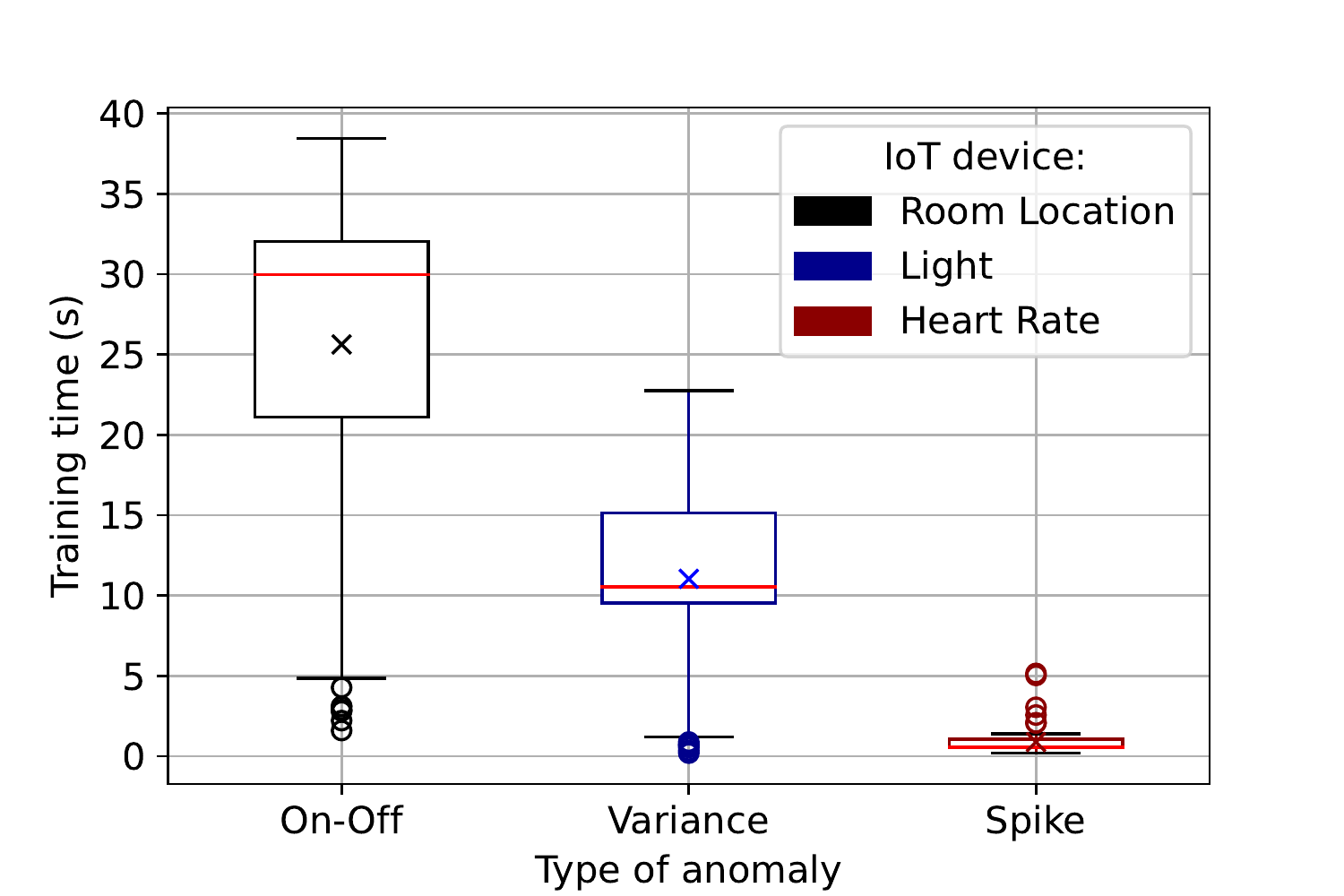}
    \caption{Anomaly detection module training time at the edge.}
    \label{fig:inference_bx}
\end{figure}

Figure~\ref{fig:inference_bx} shows the average running time
for training the neural network models on the selected edge device
(Raspberry Pi 4) with different types of anomalies. The average training time is 26 seconds for the On-Off anomaly, 11 seconds for the variance anomaly, and 0.88 seconds for the spike anomaly, demonstrating that training at the edge is feasible.

\noindent \textbf{Alternative Processing Models.}
We explore the possibility of using alternative ML models such as 
Convolutional Neural Networks (CNN)  and K-Nearest Neighbor
(KNN). Both model achieve accuracies that are significantly smaller than the DNN.

\section{Discussion}
\label{discussion}
\subsection{Limitations}
Our methodology has some limitations.

\noindent \textbf{Model degradation.} Although we show that the model retains reliability over a week-long time period, we can assume that there will be a point in time when the model's accuracy will degrade. It would be beneficial to test the accuracy degradation over a larger timescale in order to understand how often the model weights must be updated to retain accuracy.

\noindent \textbf{Scalability.} We demonstrate effective identification over a dataset of 44 patients with a relatively simple, 2-layer neural network. We do not yet understand if this model architecture will scale to a larger set of patients. It would also be useful to understand if the model could be used to accurately predict the presence of other anomalies.

\subsection{Future Directions}

One of the possible solutions that we would like to investigate in the future is to expand PRISM in order to enable privacy-preserving device behavior analytic gathering through Federated Learning (FL)~\cite{mo2021ppfl,zhao2020privacy}.
We will implement the privacy-preserving Federated Learning framework on the home gateway. This will enable the collection of anonymized and aggregated behavioral data from various IoT devices running locally while personalizing models to the usage trends of each household.

\subsection{Ethical Consideration}

We fully respected the ethical guidelines defined by our affiliated organization, and we received approval from our University’s ethical committee. Moreover, all the data used and presented in this research has been anonymized.
\section{Related Work}
\label{related}
In the last decades, a vast number of machine learning-based network monitoring and anomaly classification techniques, both
in a distributed and centralized manner, have been explored~\cite{ortiz2019devicemien}. However, not all methods are suitable for healthcare IoT, and they are not validated on real in-home monitoring datasets. 
Compared with cloud-based systems, edge-based systems~\cite{Shi2016} keep generated data on edge devices and conduct computational tasks locally. Modern edge devices now are capable of conducting advanced tasks such as training DNN models and using them for local inference~\cite{Servia2018, Chen2019}. Combining with IoT techniques, many edge-based systems have been proposed for smart healthcare~\cite{pace2018edge, Cao2015, Queralta2019, Uddin2019}, which is one of the key applications of edge computing.
Compared with existing systems, we focus on the anomaly detection task in smart healthcare and test our system with long-term and real-world data.

IoT security management is rapidly transforming from cloud-based solutions to hybrid and edge-based solutions~\cite{Zhang2019} due to privacy issues. 
Many edge-based security management solutions have been proposed in both commercial and industrial IoT. Thompson~et~al.~\cite{Thompson2021} propose an edge-based solution for rapid IoT device identification using DNS traffic. Similarly, there are other edge-based solutions using network traffic for the detection of non-essential network connections~\cite{mandalari2021blocking}, for malware detection~\cite{Kim2022}, and for intrusion detection~\cite{Ren2018}.  
These existing systems, however,  rely on analysis of network traffic, which is difficult to access especially in commercial IoT systems. In comparison, PRISM utilizes application layer data, which makes it more feasible in commercial IoT systems.
\section{Conclusion}
\label{conclusion}

Healthcare IoT devices are already very popular, and their usage is expected to grow rapidly, particularly for in-home monitoring. There is a need for a system for experimenting with IoT anomalies at the edge. 
In our study, we found that many IoT devices have a clear usage pattern and it is possible to detect such device anomalies.

In this paper, we proposed PRISM, a system enhancing privacy preserving analytics for in-home monitoring IoT devices and experimenting with anomalies. 
We trained and evaluated different ML models for IoT device anomaly detection using data collected in a clinical study by the UK Dementia Research Institute to improve care for PLWD. 

PRISM allows to inject customized anomalies and it is able to detect them with 99\% accuracy within a 24-hour training window. We demonstrate that the accuracy decreases with the training window size.
We also showed that a model trained on one patient performs poorly when tested on a different patient even from the same anomaly, indicating the need for continuous, local, and personalized retraining.

To address these issues, we evaluated PRISM at the edge using a representative edge device (Raspberry Pi 4). We showed that it is feasible to train the model at the edge with training times as low as 0.88 seconds. 

\section*{Acknowledgements}

This work was supported by the EPSRC Open Plus Fellowship (EP/W005271/1) and the EPSRC PETRAS (EP/S035362/1).
\bibliographystyle{IEEEtran}
\bibliography{main}

\begin{thebibliography}{10}
\providecommand{\url}[1]{#1}
\csname url@samestyle\endcsname
\providecommand{\newblock}{\relax}
\providecommand{\bibinfo}[2]{#2}
\providecommand{\BIBentrySTDinterwordspacing}{\spaceskip=0pt\relax}
\providecommand{\BIBentryALTinterwordstretchfactor}{4}
\providecommand{\BIBentryALTinterwordspacing}{\spaceskip=\fontdimen2\font plus
\BIBentryALTinterwordstretchfactor\fontdimen3\font minus
  \fontdimen4\font\relax}
\providecommand{\BIBforeignlanguage}[2]{{%
\expandafter\ifx\csname l@#1\endcsname\relax
\typeout{** WARNING: IEEEtran.bst: No hyphenation pattern has been}%
\typeout{** loaded for the language `#1'. Using the pattern for}%
\typeout{** the default language instead.}%
\else
\language=\csname l@#1\endcsname
\fi
#2}}
\providecommand{\BIBdecl}{\relax}
\BIBdecl

\bibitem{Habibzadeh2020}
H.~Habibzadeh, K.~Dinesh, O.~Rajabi~Shishvan, A.~Boggio-Dandry, G.~Sharma, and
  T.~Soyata, ``A survey of healthcare internet of things (hiot): A clinical
  perspective,'' \emph{IEEE Internet of Things Journal}, vol.~7, no.~1, pp.
  53--71, 2020.

\bibitem{kashani2021systematic}
M.~H. Kashani, M.~Madanipour, M.~Nikravan, P.~Asghari, and E.~Mahdipour, ``A
  systematic review of iot in healthcare: Applications, techniques, and
  trends,'' \emph{Journal of Network and Computer Applications}, vol. 192, p.
  103164, 2021.

\bibitem{ray2019systematic}
P.~P. Ray, D.~Dash, and D.~De, ``A systematic review and implementation of
  iot-based pervasive sensor-enabled tracking system for dementia patients,''
  \emph{Journal of medical systems}, vol.~43, no.~9, pp. 1--21, 2019.

\bibitem{Jaiswal2017}
K.~Jaiswal, S.~Sobhanayak, B.~K. Mohanta, and D.~Jena, ``Iot-cloud based
  framework for patient's data collection in smart healthcare system using
  raspberry-pi,'' in \emph{2017 International Conference on Electrical and
  Computing Technologies and Applications (ICECTA)}, 2017, pp. 1--4.

\bibitem{Meingast}
M.~Meingast, T.~Roosta, and S.~Sastry, ``Security and privacy issues with
  health care information technology,'' in \emph{2006 International Conference
  of the IEEE Engineering in Medicine and Biology Society}, 2006, pp.
  5453--5458.

\bibitem{mandalari2021blocking}
A.~M. Mandalari, D.~J. Dubois, R.~Kolcun, M.~T. Paracha, H.~Haddadi, and
  D.~Choffnes, ``Blocking without breaking: Identification and mitigation of
  non-essential iot traffic,'' \emph{Proceedings on Privacy Enhancing
  Technologies}, vol.~4, 2021.

\bibitem{velasco2018optimum}
D.~Velasco-Montero, J.~Fern{\'a}ndez-Berni, R.~Carmona-Gal{\'a}n, and
  {\'A}.~Rodr{\'\i}guez-V{\'a}zquez, ``Optimum selection of dnn model and
  framework for edge inference,'' \emph{IEEE Access}, vol.~6, pp.
  51\,680--51\,692, 2018.

\bibitem{zhao2018privacy}
J.~Zhao, R.~Mortier, J.~Crowcroft, and L.~Wang, ``Privacy-preserving machine
  learning based data analytics on edge devices,'' in \emph{Proceedings of the
  2018 AAAI/ACM Conference on AI, Ethics, and Society}, 2018, pp. 341--346.

\bibitem{osia2018private}
S.~A. Osia, A.~S. Shamsabadi, A.~Taheri, H.~R. Rabiee, and H.~Haddadi,
  ``Private and scalable personal data analytics using hybrid edge-to-cloud
  deep learning,'' \emph{Computer}, vol.~51, no.~5, pp. 42--49, 2018.

\bibitem{pace2018edge}
P.~Pace, G.~Aloi, R.~Gravina, G.~Caliciuri, G.~Fortino, and A.~Liotta, ``An
  edge-based architecture to support efficient applications for healthcare
  industry 4.0,'' \emph{IEEE Transactions on Industrial Informatics}, vol.~15,
  no.~1, pp. 481--489, 2018.

\bibitem{wu2021edge}
F.~Wu, C.~Qiu, T.~Wu, and M.~R. Yuce, ``Edge-based hybrid system implementation
  for long-range safety and healthcare iot applications,'' \emph{IEEE Internet
  of Things Journal}, vol.~8, no.~12, pp. 9970--9980, 2021.

\bibitem{kolcunrevisiting}
R.~Kolcun, D.~A. Popescu, V.~Safronov, P.~Yadav, A.~M. Mandalari, R.~Mortier,
  and H.~Haddadi, ``Revisiting iot device identification,'' \emph{TMA
  Conference}, 2022.

\bibitem{palermo2021designing}
S.~Kouchaki, F.~Palermo, H.~Li, A.~Capstick, N.~Fletcher-Lloyd, Y.~Zhao,
  R.~Nilforooshan, D.~Sharp, and P.~Barnaghi, ``Designing a clinically
  applicable deep recurrent model to identify neuropsychiatric symptoms in
  people living with dementia using in-home monitoring data,'' in \emph{35th
  Conference on Neural Information Processing Systems (NeurIPS 2021)}, 2021.

\bibitem{chen1999rapid}
C.~P. Chen and J.~Z. Wan, ``A rapid learning and dynamic stepwise updating
  algorithm for flat neural networks and the application to time-series
  prediction,'' \emph{IEEE Transactions on Systems, Man, and Cybernetics, Part
  B (Cybernetics)}, vol.~29, no.~1, pp. 62--72, 1999.

\bibitem{hunter2012selection}
D.~Hunter, H.~Yu, M.~S. Pukish~III, J.~Kolbusz, and B.~M. Wilamowski,
  ``Selection of proper neural network sizes and architectures—a comparative
  study,'' \emph{IEEE Transactions on Industrial Informatics}, vol.~8, no.~2,
  pp. 228--240, 2012.

\bibitem{moulines2011non}
E.~Moulines and F.~Bach, ``Non-asymptotic analysis of stochastic approximation
  algorithms for machine learning,'' \emph{Advances in neural information
  processing systems}, vol.~24, 2011.

\bibitem{mo2021ppfl}
F.~Mo, H.~Haddadi, K.~Katevas, E.~Marin, D.~Perino, and N.~Kourtellis, ``Ppfl:
  privacy-preserving federated learning with trusted execution environments,''
  in \emph{Proceedings of the 19th Annual International Conference on Mobile
  Systems, Applications, and Services}, 2021, pp. 94--108.

\bibitem{zhao2020privacy}
Y.~Zhao, H.~Haddadi, S.~Skillman, S.~Enshaeifar, and P.~Barnaghi,
  ``Privacy-preserving activity and health monitoring on databox,'' in
  \emph{Proceedings of the Third ACM International Workshop on Edge Systems,
  Analytics and Networking}, 2020, pp. 49--54.

\bibitem{ortiz2019devicemien}
J.~Ortiz, C.~Crawford, and F.~Le, ``Devicemien: network device behavior
  modeling for identifying unknown iot devices,'' in \emph{Proceedings of the
  International Conference on Internet of Things Design and Implementation},
  2019, pp. 106--117.

\bibitem{Shi2016}
W.~Shi, J.~Cao, Q.~Zhang, Y.~Li, and L.~Xu, ``Edge computing: Vision and
  challenges,'' \emph{IEEE Internet of Things Journal}, vol.~3, no.~5, pp.
  637--646, 2016.

\bibitem{Servia2018}
S.~Servia-Rodríguez, L.~Wang, J.~R. Zhao, R.~Mortier, and H.~Haddadi,
  ``Privacy-preserving personal model training,'' in \emph{2018 IEEE/ACM Third
  International Conference on Internet-of-Things Design and Implementation
  (IoTDI)}, 2018, pp. 153--164.

\bibitem{Chen2019}
J.~Chen and X.~Ran, ``Deep learning with edge computing: A review,''
  \emph{Proceedings of the IEEE}, vol. 107, no.~8, pp. 1655--1674, 2019.

\bibitem{Cao2015}
\BIBentryALTinterwordspacing
Y.~Cao, P.~Hou, D.~Brown, J.~Wang, and S.~Chen, ``Distributed analytics and
  edge intelligence: Pervasive health monitoring at the era of fog computing,''
  in \emph{Proceedings of the 2015 Workshop on Mobile Big Data}, ser. Mobidata
  '15.\hskip 1em plus 0.5em minus 0.4em\relax New York, NY, USA: Association
  for Computing Machinery, 2015, p. 43–48. [Online]. Available:
  \url{https://doi.org/10.1145/2757384.2757398}
\BIBentrySTDinterwordspacing

\bibitem{Queralta2019}
J.~P. Queralta, T.~N. Gia, H.~Tenhunen, and T.~Westerlund, ``Edge-ai in
  lora-based health monitoring: Fall detection system with fog computing and
  lstm recurrent neural networks,'' in \emph{2019 42nd International Conference
  on Telecommunications and Signal Processing (TSP)}, 2019, pp. 601--604.

\bibitem{Uddin2019}
\BIBentryALTinterwordspacing
M.~Z. Uddin, ``A wearable sensor-based activity prediction system to facilitate
  edge computing in smart healthcare system,'' \emph{Journal of Parallel and
  Distributed Computing}, vol. 123, pp. 46--53, 2019. [Online]. Available:
  \url{https://www.sciencedirect.com/science/article/pii/S0743731518306270}
\BIBentrySTDinterwordspacing

\bibitem{Zhang2019}
J.~Zhang, H.~Chen, L.~Gong, J.~Cao, and Z.~Gu, ``The current research of iot
  security,'' in \emph{Proceedings of the 2019 IEEE Fourth International
  Conference on Data Science in Cyberspace (DSC)}, 2019, pp. 346--353.

\bibitem{Thompson2021}
O.~Thompson, A.~M. Mandalari, and H.~Haddadi, ``Rapid iot device identification
  at the edge,'' in \emph{Proceedings of the 2nd ACM International Workshop on
  Distributed Machine Learning}, 2021, pp. 22--28.

\bibitem{Kim2022}
\BIBentryALTinterwordspacing
H.-m. Kim and K.-h. Lee, ``Iiot malware detection using edge computing and deep
  learning for cybersecurity in smart factories,'' \emph{Applied Sciences},
  vol.~12, no.~15, 2022. [Online]. Available:
  \url{https://www.mdpi.com/2076-3417/12/15/7679}
\BIBentrySTDinterwordspacing

\bibitem{Ren2018}
W.~Ren, T.~Yardley, and K.~Nahrstedt, ``Edmand: Edge-based multi-level anomaly
  detection for scada networks,'' in \emph{2018 IEEE International Conference
  on Communications, Control, and Computing Technologies for Smart Grids
  (SmartGridComm)}, 2018, pp. 1--7.

\end{thebibliography}

\end{document}